\documentclass[12pt]{article}

\usepackage{amsmath}
\usepackage{amssymb}
\usepackage{dsfont} 

\numberwithin{equation}{section}

\usepackage{graphicx}

\usepackage[dvipsnames]{xcolor}
\usepackage{color}

\newcommand{\be}{\begin{equation}}
\newcommand{\ee}{\end{equation}}
\newcommand{\bel}[1]{\begin{equation}\label{#1}}
\newcommand{\bea}{\begin{eqnarray}}
\newcommand{\eea}{\end{eqnarray}}
\newcommand{\balign}{\begin{align}}
\newcommand{\ealign}{\end{align}}
\newcommand{\ba}{\begin{array}}
\newcommand{\ea}{\end{array}}
\newcommand{\bfig}{\begin{figure}}
\newcommand{\efig}{\end{figure}}

\newcommand{\eref}[1]{(\ref{#1})}

\newcommand{\Fref}[1]{Figure~\ref{#1}}

\allowdisplaybreaks

\newcommand{\exval}[1]{\mbox{$\langle \, {#1}\, \rangle$}}

\newcommand{\bfm}{\mathbf{m}}
\newcommand{\bfs}{\mathbf{s}}
\newcommand{\bfx}{\mathbf{x}}

\newcommand{\rmd}{\mathrm{d}}
\newcommand{\rme}{\mathrm{e}}

\newcommand{\ddt}{\frac{\rmd}{\rmd t}}

\newcommand{\half}{\frac{1}{2}}

\newcommand{\bzeta}{\boldsymbol{\zeta}}
\newcommand{\bfeta}{\boldsymbol{\eta}}

\newcommand{\PPi}{PP\textsubscript{i} }

\begin{document}

\title{RNA polymerase interactions and elongation rate}
\author{V. Belitsky$^{1}$ 
\and G.M.~Sch\"utz$^{2}$
}
\date{}
\maketitle

{\small
\noindent $^{~1}$Instituto de Matem\'atica e Est\'atistica,
Universidade de S\~ao Paulo, Rua do Mat\~ao, 1010, CEP 05508-090,
S\~ao Paulo - SP, Brazil\\
\noindent Email: belitsky@ime.usp.br 

\smallskip
\noindent $^{~2}$Institute of Complex Systems II,
Theoretical Soft Matter and Biophysics,
Forschungszentrum J\"ulich, 52425 J\"ulich, Germany\\
\noindent Email: g.schuetz@fz-juelich.de
}

\begin{abstract}

We show that non-steric molecular interactions between RNA polymerase (RNAP) motors 
that move simultaneously on the same DNA track determine strongly the kinetics of 
transcription elongation. 
With a focus on the role of collisions and cooperation, we introduce a stochastic model 
that allows for the exact analytical computation of the stationary properties of 
transcription elongation as a function 
of RNAP density, their interaction strength, nucleoside triphosphate concentration, and rate of pyrophosphate 
release. Cooperative pushing, i.e., an enhancement of the 
average RNAP velocity and elongation rate, arises due to stochastic pushing. This cooperative effect cannot be explained by steric hindrance alone but requires a sufficiently 
strong molecular repulsion. It
disappears beyond a critical RNAP density, above which jamming due to collisions takes over. 
For strong stochastic blocking the
cooperative pushing is suppressed at low RNAP densities, but a reappears at higher 
densities.
%

\end{abstract}

\newpage

\section{Introduction}

RNA polymerase (RNAP) is a molecular motor that transcribes the information coded in the 
base pair sequence of DNA into an RNA. The process is initiated at a promoter sequence on the DNA.
Stepping along the base pairs of the DNA, the RNAP forms the transcription elongation complex (TEC) 
which polymerizes the monomeric subunits of the RNA by the addition of nucleotides, as determined 
by the corresponding sequence on the template DNA \cite{Albe13,Bai06}. To this end,
the RNAP locally creates the so-called transcription bubble by unzipping the two DNA strands as it 
progresses. The elongation process ends when the TEC reaches the termination sequence. Then the 
nascent RNA is released by the RNAP
and serves as a template for translation into proteins (in the case of messenger RNA) or facilitates 
the formation of RNA-protein complexes such as ribosomes. Thus RNAPs 
play a central role in gene expression and also as therapeutic drug targets \cite{Ma16}.

Here we focus on the kinetics of the elongation stage where each successful 
addition of ribonucleotides to the growing RNA transcript is accompanied by
a biased random walk of the RNAP along the DNA with a step length of one base pair
for each translocation \cite{Guth99,Omao11}. 
Our main interest is the impact of interactions between RNAP on the average speed
of an individual RNAP which determines the overall rate of elongation.

The intrinsically stochastic dynamics of a single RNAP has been studied in great detail. 
The major and generic features of the multi-step mechano-chemical pathway of individual RNAP motors 
during each elongation step are:  (1) Nucleoside triphosphate  (NTP) binding at a catalytic site 
within the transcription bubble, (2) NTP hydrolysis, (3) Release of pyrophosphate  (PP\textsubscript{i}, 
one of the products of hydrolysis), (4) Forward step of the RNAP along the DNA template by 
one base pair (bp) \cite{Bai06,Wang98}. Recently, the mechanism of NTP binding to the active site 
\cite{Wu17} and PP\textsubscript{i} release \cite{Da15} were studied with atomistic simulations 
in great detail for T7 RNA Polymerase. Also recently, the role of trigger loop folding-unfolding 
on the RNAP translocation was elucidated for the RNAP of Escherichia coli \cite{Meji15}. For
a recent more detailed description, see \cite{Zhan16}.

If more than one RNAP molecule initiates from the same promoter one cannot ignore their mutual 
interactions. On the one hand, pausing RNAP may block the advancement of trailing RNAP and thus 
induce ``traffic jams'' \cite{Klum08,Trip08} and phase transitions in the rate of elongation 
\cite{Trip09}. On the other hand, by the same token the interaction has been 
demonstrated to be cooperative: Trailing RNAP strongly prevent backtracking, and even 
``push'' the leading RNAP out of pause sites \cite{Epsh03a,Epsh03b,Saek09,Jin10,Ehre10,Galb11,Klum11}. 
Thus one is faced with the apparently paradoxical picture that the appearance of traffic jam would
suggest a reduction of the average rate of elongation, while cooperative behaviour indicates 
an enhancement. 

How RNAP interactions affect the rate of elongation is not known 
quantitatively nor is their precise nature. Most modelling approaches to molecular motor traffic
that are based on generalizations of the asymmetric simple exclusion process (ASEP),
a generic Markovian lattice gas model, assume only a hard core repulsion corresponding to 
pure steric hindrance on contact. This approach successfully captures the traffic jam
phenomenon and other kinetic properties of the transcription process \cite{Scha10}. However,
this interaction does not explain pushing as the rate of forward steps along the DNA in such
models does not depend on the presence of motors in the backward direction.

This shortcoming is addressed in a number of recent works. Galburt et al. \cite{Galb11} employed transition rate
theory to determine rates of translocation by postulating a phenomenological free-energy landscape
for two neighbouring RNAP interacting via a hard-core potential. They concluded that pushing occurs
only in cases where the transition state of the active particle is relatively early during a step.
Costa et al. \cite{Cost13} introduced a model
that generalizes the stochastic sequence-dependent model of Bai et al. \cite{Bai04,Bai07} such
that collisions between elongating RNAP modify their elongation rates. For the 
sequences that were investigated by numerical simulation an acceleration of transcription 
up to 48\% was observed when collisions are allowed. Teimouri et al \cite{Teim15}
consider a generic Markovian model for molecular motors where the pushing and collisions are
explicitly put into the model by rates that are different from the single-motor step rate
when a motor attempts to move close to a paused motors or when it 
moves away from it. With numerical simulation and analytical mean field approximation it is 
shown that there is an optimal repulsive interaction strength that leads to a 
maximal particle flux, which in the context of our discussion translates into an optimal
elongation rate. In this model the mechano-chemical cycle that RNAP undergo during
translocation is ignored. Heberling et al. \cite{Hebe16} argue that
the torque produced by RNAP motion on helically twisted DNA leads to fewer
collisions and thus explains transcription enhancement. This hypothesis is supported
by Monte-Carlo simulation of a generalized ASEP in which the rate of translocation 
depends on the torque between the polymerase and its closest two neighboring polymerases. 
The amount of torque is, in turn, the result of the relative motion of RNAPs on the DNA strand.
However, the notion of transcription enhancement due to a reduced number
of collisions is at odds with the pushing picture which relies on collisions of the trailing
RNAP with a pausing RNAP. 

Here we take a different approach. We propose a stationary distribution of the chemical states and 
the relative distances between RNAPs along the template DNA. From this distribution we determine 
Markovian stochastic transition mechanisms that on the one hand are compatible with stationarity and on the other hand
take into account the main processes in the mechano-chemical RNAP stepping cycle as well as not only steric 
hard-core repulsion, but also a short-range term that captures the interaction when two RNAP are very close to 
each other. This approach enables us to make \emph{exact} analytical computations for collective stationary 
quantities rather than having to rely on numerical simulation or analytical mean-field theory 
which do not allow for any rigorous estimate of the approximation error. 

It will transpire that hard-core repulsion alone does not allow to capture any pushing phenomenon.
However, the presence of an additional short-range repulsion turns out to
lead to \emph{stochastic pushing} (which is not an input into our model) and thus explains it as a 
general consequence of short-range repulsion between neighbouring RNAP moving stochastically along the DNA track.
The effect of the stochastic pushing on the collective average motor velocity and hence on the
average elongation rate depends on the average density of interacting RNAP 
along the DNA. At sufficiently low densities the cooperative effect of pushing prevails over 
blocking and leads to an enhanced RNAP speed, compared to the average speed of an 
isolated single motor. Beyond a critical density, however,
the average velocity starts to decrease, even though the average elongation rates generated 
by the collective action of all RNAP still increases with density. At a second higher critical density 
blocking prevails and also the average elongation rate decreases with further increasing density.
Thus our approach accounts both for cooperative pushing and jamming. Interestingly,
for sufficiently strong repulsion the average elongation rate can exhibit two maxima at different 
RNAP densities. Moreover,
the model makes analytical predictions about how the motor velocity and elongation rate
change when the concentration of NTP and the rate of PP\textsubscript{i} release vary.

In the next section we formulate the mathematical model and review some of the theoretical and empirical
background that motivates it. In the third section we present and discuss our results. First we focus on minimal-range 
interaction and present exact analytic expressions for the RNAP average velocity and transcription 
rate as functions of the RNAP density and interaction strength. Then we extend the discussion to
an extended interaction range of the microscopic processes without impact on the stationary 
distribution. In the concluding section we summarize the gist of our approach
and the main findings. The mathematical derivations of the results are presented in an Appendix.

\section{Methods}

We address the questions detailed in the introduction with a stochastic model for 
interacting RNAP that we introduce here. The model is a Markov process whose exact 
stationary properties, in particular, the average rate of transcription elongation, 
can be computed analytically as function of the model parameters.
First we describe the model for the stochastic dynamics of an isolated RNAP, following \cite{Wang98}. 
In a second step we introduce interactions.

\subsection{Single-RNAP dynamics}

As mentioned above, the RNAP undergoes various transformations during each elongation step,
see \cite{Zhan16} for a recent discussion. Here we do not aspire a detailed description.
In the reduced description of Wang et al \cite{Wang98}, which accounts rather well 
for various experimentally established features of the kinetics of single RNAP, and which 
is adequate as a starting point for the purposes of this work, the rate-limiting step is the 
PP\textsubscript{i} release from the catalytic site \cite{Erie92}. An RNAP carrying an RNA 
transcript of length $n$ thus appears in only two distinct polymerization states, namely 
with or without PP\textsubscript{i} bound to it. It moves forward along the DNA only after 
PP\textsubscript{i} release by one bp, corresponding to a step length $\delta=0.34nm$.

\begin{figure}[ptb]
\begin{center}
\includegraphics[scale=0.5]{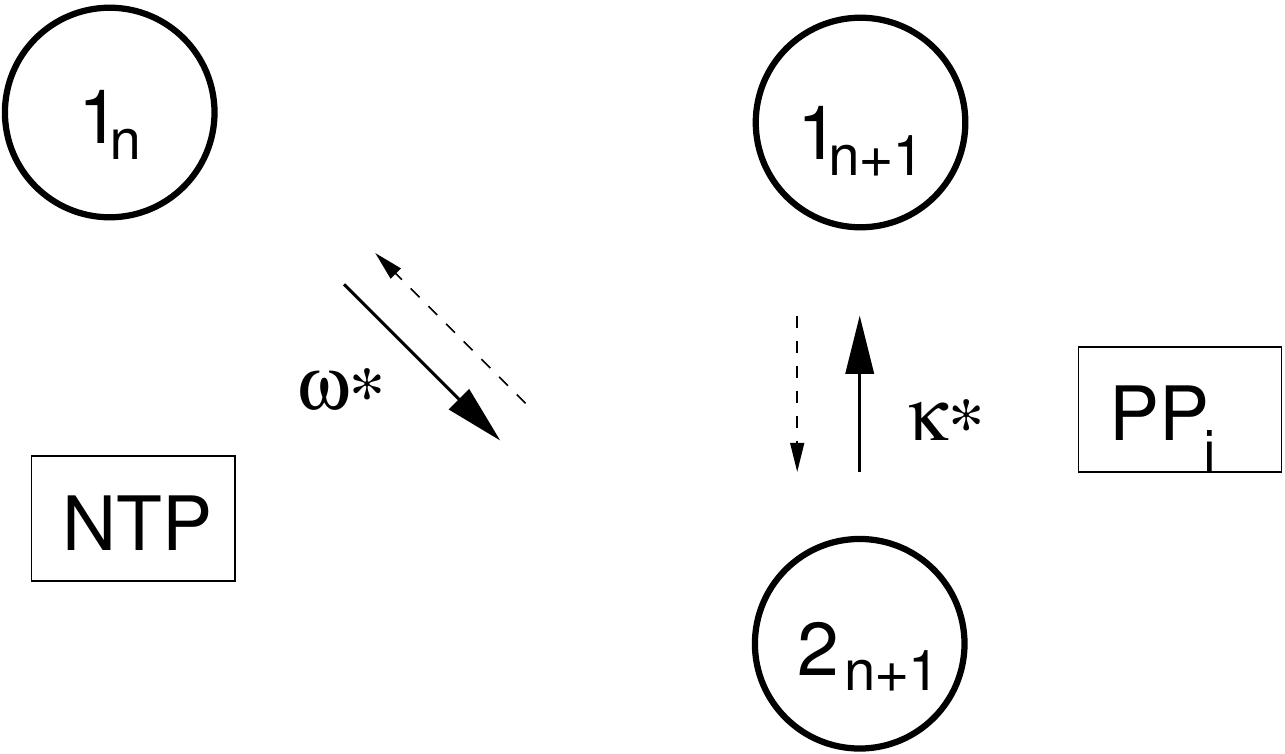}
\end{center}
\caption{
Minimal reaction scheme of RNAP translocation. The RNAP can move from base pair $n$ to $n+1$
provided it is in state $1_n$ (no pyrophospate bound to it) with an effective rate 
$\omega^{\star}$ through a series of processes involving NTP binding to the 
active site and NTP hydrolysis which results in a PP\textsubscript{i} bound state 
in the transcription elongation complex (state $2_{n+1}$). Only after PP\textsubscript{i} release with
rate $\kappa^{\star}$ (transition from state $2_{n+1}$ to state $1_{n+1}$) the RNAP can perform 
the next translocation
step. Reverse reactions (not considered here) are indicated with thin dashed arrows.
}%
\label{Fig:Reactionscheme}%
\end{figure}

It is natural to describe the state of an RNAP mathematically not in terms of the length 
$n$ of the RNA transcript attached to it, but to consider $n$ to mark the corresponding 
base pair on the template DNA. We denote the RNAP state without PP\textsubscript{i} 
by 1 and when it is bound by 2. Thus the translocation of the RNAP from base pair $n$
to $n+1$ corresponds to a directed random step on a one-dimensional lattice from site 
$n$ to $n+1$ provided the RNAP is in state 1.
We denote the rate at which this step occurs by $\omega^{\star}$. The processes that lead to 
translocation are fast by comparison to the PP\textsubscript{i} release and thus conflate
into the translocation step itself, which means that the RNAP arrives at site $n+1$ in state 2
with PP\textsubscript{i} bound. It can move again only after PP\textsubscript{i} is released,
which happens with a rate that we denote by $\kappa^{\star}$. This reduced minimal reaction 
scheme is sketched in \Fref{Fig:Reactionscheme}.

The reverse processes are possible, but occur with smaller rates \cite{Wang98} that we neglect.
For the purpose of elucidating the role of interactions we also ignore the sequence dependence
and the role of the trigger loop in the translocation mechanism.
Following \cite{Wang98} we assume that the PP\textsubscript{i} release 
rate $\kappa^{\star}$ is independent of the NTP concentration while the translocation rate 
$\omega^{\star}$ is proportional to it. In the setting considered by Wang et al. one has 
under load-free conditions
\bel{omega}
\kappa^{\star} = 31.4 s^{-1}, \quad \omega^{\star} = [NTP] (\mu M)^{-1} s^{-1} 
\ee
where $[NTP]$ is the NTP concentration. The rates $\kappa^{\star}$ and 
$\omega^{\star}$ are of similar magnitude for $[NTP]$ around $30 \mu M$. The translocation 
rate $\omega^{\star}$ is a free parameter of our model.
Under cellular conditions, RNAP transription leads to downstream 
supercoiling that generates a load force of $6 pN$ \cite{Yin95}. We ignore this
effect as it only renormalizes the translocation rate by a factor of approximately 
$\exp{(-6 pN \cdot 0.34 nm)/k_BT} \approx 0.6$.

\subsection{RNAP interactions}

We extend the two-state random walk model for single RNAP to a Markov process of interacting 
random walks. The typical size of a transcription bubble is around 15 bp \cite{Wang98,Gesz05} whereas
the TEC covers a DNA segment of up to 35 bp. We simplify the complicated geometry of the TEC
by representing it as a rod covering $\ell$ lattice sites, where $\ell$ is parameter of our model
that can be adjusted as needed.
We number the RNAPs sequentially from 1 to $N$ and denote
by $x_i$ the (integer) lattice position of the left end of the $i^{th}$ rod in the random walk model, 
which corresponds to the trailing side of the corresponding TEC. 
Then $x_i+\ell-1$ is the lattice position of the ``front'' side of the TEC. 
We say that two rods (TECs) $i$ and $i+1$ are neighbours when the front of rod 
$i$ and the left edge of rod $i+1$ occupy neighbouring lattice sites, i.e., when $x_{i+1} = x_{i}+\ell$. 
For technical reasons and since we are interested only in the elongation stage of transcription we take a lattice of 
$L$ sites with periodic boundary conditions.

\subsubsection{Short-range contribution to hard-core repulsion}

Both for RNAP and other molecular motors the modelling of steric hindrance 
by a hard-core repulsion alone has a very long history, starting with the pioneering
work of MacDonald et al \cite{MacD68} on ribosomes, see e.g. \cite{Trip08,Shaw03,Klum03,Chou04,Frey04,Dong07,Greu07,Kolo07,Chow13} 
for later developments concerning ribosomes, RNAP and other molecular motors.  
In the framework of our model this means that translocation cannot occur if the site $x_i+\ell$ next to the tip 
of the RNAP $i$ is occupied by a leading RNAP $i+1$. Thus 
traffic jams of molecular motors may form \cite{Chow13}. However, as shown below,
this approach does not allow for pushing.

Indeed, recent data suggest that the description of the free energy landscape of two
neighbouring TECs by just a hard-core interaction is inappropriate when the leading polymerase 
is unable to move forward due to nucleotide starvation. Under these conditions, the system 
exhibits elastic deformations \cite{Saek09,Galb11}. 
In order to capture this effect we introduce an interaction that 
has short-ranged contributions to hard-core repulsion. Since we do not wish to use
pushing as an input into model but \emph{derive} it analytically, we define 
processes along the lines indicated above for a single RNAP, 
but with rates that depend on the presence of nearby RNAP. These rates are given initially 
as free parameters. Then we fix mathematically rigorously relations between the proposed transition rates 
to ensure stationarity of the distribution. This will indicate whether the rates are in a range compatible
with pushing or not.

\subsubsection{Stationary distribution}

Hardcore repulsion means that configurations such that $x_{i+1} = x_i+\ell-1$
are forbidden since in such a case the front of rod $i$ and the end of rod $i+1$
would occupy the same site. Correspondingly, the stationary distribution gives
give probability 0 to such configurations. Moreover, since TECs cannot overtake
each other, the coordinates $x_i$ will stay ordered at all times, i.e., only
RNAP configurations satisfying the ordering condition $x_{i+1} \geq x_i+\ell$ 
have non-zero probability. We shall refer to such configurations as allowed 
configurations. An allowed configuration of $N$ RNAPs, which we denote by $\bfeta$, 
is thus specified by a coordinate vector $\bfx = (x_1,\dots,x_N)$ with ordered integer 
coordinates and a state vector $\bfs = (s_1,\dots,s_N)$ with state variables $s_i\in\{1,2\}$.

For the stationary distribution of the interacting RNAPs we assume for the interaction part 
of $N$ RNAPs at positions $\bfx=(x_1,\dots,x_N)$ in addition to hardcore repulsion
an effective short range interaction energy of the form
\bel{energy}
U(\bfx) = J \sum_{i=1}^N \delta^L_{x_{i+1},x_i+\ell}.
\ee
Here $\delta^L$ denotes the Kronecker symbol with arguments understood modulo $L$
due to periodic boundary conditions. Positive $J$ corresponds to repulsion. 

We denote by $N^\alpha$ the fluctuating number of RNAPs in state $\alpha\in\{1,2\}$,
so that $N=N^1+N^2$. We also define the excess $B(\bfs)=N^1-N^2$. The stationary distribution 
for allowed configurations thus takes the form
\bel{sd} 
\pi^\ast(\bfeta) = \frac{1}{Z}  \pi(\bfeta)
\ee
with the Boltzmann weights
\bel{Bw}
\pi(\bfeta) = 
\exp{\left[- \frac{1}{k_BT} \left(U + \lambda B\right)\right]} 
\ee
and the partition function
\bel{pf} 
Z = \sum_{\bfeta} \pi(\bfeta).
\ee
The chemical potential $\lambda$ is a Lagrange multiplier that takes care of the fluctuations
in the excess 
\bel{excess}
B(\bfs) = \sum_{i=1}^N (3-2s_i)
\ee
due to the interplay of NTP hydrolysis and PP\textsubscript{i} release. 

For later use we introduce the parameters
\bel{xydef}
x = \rme^{\frac{2\lambda}{k_BT}}, \quad y = \rme^{\frac{J}{k_BT}}
\ee
to work with, instead of $\lambda$ and $J$. Positive $\lambda$ (i.e., $x>1$) corresponds to
an excess of RNAP in state 1. Repulsive interaction corresponds to $y>1$.

\subsubsection{Kinetic assumptions}

The rate-limiting processes that occur during transcription elongation must be consistent
with the stationarity of the distribution \eref{sd}. For the
the kinetics of a single RNAP model we adopt the results of \cite{Wang98} for the
temperature and concentration dependence
of the translocation rate $\omega^{\star}$ appearing in \eref{omega} as well as a constant value
for $\kappa^{\star}$. However, as mentioned above, in the interacting case we allow these 
rates to depend
parametrically on the presence of neighbouring RNAP.
We consider separately 
two different forms of a kinetic short-range interaction which influence the rates of translocation 
and PP\textsubscript{i} release.

\paragraph{A) Minimal kinetic interaction range}
The rate for a move by one step 
of rod $i$ covering sites from $x_{i}$ to $x_{i}+\ell-1$ 
depends on the occupation of the directly neighboring sites $x_{i-1}$ and $x_{i+\ell}$. As in the single-RNAP
case the rod needs to be in state 1 to move and arrives at the target site in state 2. 

From the hard-core repulsion we require that an RNAP can move forward by 
one site, if demanded by its own mechanochemistry, provided 
the target site $x_{i+\ell}$ is not already covered. 
Thus the configuration-dependent translocation rate for rod $i$ from position $x_i$ to position $x_{i}+1$ 
is of the form
\bel{jrA}
\omega_i(\bfeta) = \omega^{\star} \delta_{s_i,1} \left(1 + d^{1\star} \delta_{x_{i-1}+\ell , x_{i}} \right)
\left(1-\delta_{x_{i}+\ell, x_{i+1}}\right)
\ee
with ``bare'' rate $\omega^{\star}$ when both neighbouring sites 
$x_{i}-1$ and $x_{i+\ell}$ are empty and a correction given by the phenomenological dimensionless interaction parameter
$d^{1\star} \geq -1$ (to ensure positivity of the rate) when rods $i-1$ and $i$ are neighbours. The overall factor 
$\left(1-\delta_{x_{i+\ell},x_{i+1}}\right)$ is zero when rods $i$ and $i+1$ are neighbours 
and thus expresses hard-core repulsion.

Likewise, we parametrize the rate for PP\textsubscript{i} release as
\bel{rrA}
\kappa_i(\bfeta) = \kappa^{\star} \delta_{s_i,2} \left(1 + f^{1\star} \delta_{x_{i-1}+\ell ,x_{i}} 
+ f^{\star1} \delta_{x_{i}+\ell ,x_{i+1}}  + f^{1\star1} 
\delta_{x_{i-1}+\ell, x_{i}} \delta_{x_{i}+\ell, x_{i+1}} \right)
\ee
with the bare release rate $\kappa^{\star}$ and interaction terms coming from the potential
presence of neighbouring rods.

The master equation for the probability $P(\bfeta,t)$ of finding the rods at time $t$ in a configuration
$\bfeta$ thus reads
\bel{me1}
\ddt P(\bfeta,t) = \sum_{i=1}^N \left[\omega_i(\bfeta_{tl}^{i}) P(\bfeta_{tl}^{i},t) 
+ \kappa_i(\bfeta_{rel}^{i}) P(\bfeta_{rel}^{i},t) - (\omega_i(\bfeta)+\kappa_i(\bfeta))P(\bfeta,t)\right]
\ee
where $\bfeta_{tl}^{i}$ is the configuration that leads to $\bfeta$ before a translocation of RNAP $i$ 
(i.e., with coordinate $x^{tl}_{i}=x_i-1$ and state $s^{tl}_{i}=3-s_i$) and $\bfeta_{rel}^{i}$
is the configuration $\bfeta$ before PP\textsubscript{i} release at RNAP $i$ (i.e., with
coordinate $x^{rel}_{i}=x_i$ and state $s^{rel}_{i}=3-s_i$. Due to periodicity, the positions
$x_i$ of the rods are counted modulo $L$ and labels $i$ are counted modulo $N$.

\paragraph{B) Extended kinetic interaction range} As we will see, the form \eref{sd} of the
stationary distribution does not uniquely fix the rates for translocation and PP\textsubscript{i} 
release. In particular, one may explore a dependence of these rates over distances of neighbouring
rods of more than one site. 

For the configuration-dependent translocation rate for rod $i$ we assume 
a dependence on whether the
configuration after the jump leads to a new neighbouring pair of rods or not, as this changes
the effective energy \eref{energy} and hence the Boltzmann weight \eref{Bw}
of the configuration in the distribution \eref{sd}. This yields
\bea
\label{jrB}
\omega_i(\bfeta) & = & 
\omega^{\star} \delta_{s_i,1} \left(1 + d^{1\star} \delta_{x_{i-1}+\ell , x_{i}} 
+ d^{\star1} \delta_{x_{i+1}, x_{i}+\ell + 1} \right)
\left(1-\delta_{x_{i+1}, x_{i}+\ell}\right) 
\eea
with a further dimensionless interaction parameter $d^{\star1}$.

For the rate of PP\textsubscript{i} release, which does no lead to any change
in the effective energy \eref{energy}, we allow for a dependence on the presence 
neighboring
RNAP of the form
\bea
\label{rrB}
\kappa_i(\bfeta) & = & \kappa^{\star} \delta_{s_i,2} \left[1 + f^{1\star} \delta_{x_{i-1}+\ell ,x_{i}} 
+ f^{\star1} \delta_{x_{i}+\ell ,x_{i+1}}  + f^{1\star1} 
\delta_{x_{i-1}+\ell ,x_{i}} \delta_{x_{i}+\ell ,x_{i+1}} \right. \nonumber \\
& & \left. + f^{10\star} (1-\delta_{x_{i-1}+\ell ,x_{i}}) \delta_{x_{i-1}+\ell+1 ,x_{i}} 
+ f^{\star01} (1-\delta_{x_{i}+\ell ,x_{i+1}})  \delta_{x_{i+1}, x_{i}+\ell +1}\right]  .
\eea

The master equation for the extended process is of the same form as the master equation \eref{me1} 
since the allowed 
transitions are the same, the only difference being their rate. We do not specify the interactions 
that lead to the parameters appearing in the rates \eref{jrA},\eref{rrA}, \eref{jrB}, and \eref{rrB}
since we do not have sufficient experimental knowledge of the underlying processes 
on molecular level. We stress that one cannot impose detailed balance since the system is not 
in
thermal equilibrium. We can just state that for a stationary distribution the left hand side 
of the many-particle master equation \eref{me1} is equal to 0, by definition of
stationarity. Thus the condition of stationary of the proposed distribution \eref{sd}
determines the parameters that enter the interaction the parameters appearing 
in the rates \eref{jrA},\eref{rrA}, \eref{jrB}, and \eref{rrB}.

\section{Results and Discussion}

Due to the stochastic dynamics all quantities of interest are fluctuating. However, under high growth
conditions a large number of RNAP transcribe simultaneously and fluctuations get washed out.
Hence in this work our main concern are stationary averages. These are not averages over
different cells where various external factors may lead to variations, but averages over the 
fluctuations inherent to RNAP transcription from the same DNA template. We denote the
average density of RNAP by $\rho$ by which we mean the average number of RNAP on a
DNA sequence of $L$ bp. We also recall the definitions \eref{xydef} that will be used extensively.

\subsection{Stationarity condition}

As pointed out above, given the stationary distribution \eref{sd}, the transition rates 
$\omega_i(\bfeta)$ and $\kappa_i(\bfeta)$ cannot be 
chosen freely. We stress that, in fact, it is not even {\it a priori} clear whether a process
exists for which the distribution \eref{sd} is stationary and which has rates 
of the form \eref{jrA} and \eref{rrA} 
or \eref{jrB} and \eref{rrB}. It is therefore rather non-trivial that, as shown in the Appendix, 
the requirement of stationarity of the probability distribution \eref{sd} fixes in a non-trivial
fashion the parameters 
$x,y$ \eref{xydef} and the rates \eref{jrB} and \eref{rrB}, up to
four free parameters $\kappa^{\star}, \omega^{\star}, d^{1\star}, d^{\star1}$. 
In terms of these parameters one obtains the relations
\bea
\label{xB}
& & x = \frac{\omega^{\star}}{\kappa^{\star}} \\
\label{yB}
& & y = \frac{1+d^{1\star}}{1+d^{\star1}} \\
\label{f1sB}
& & f^{1\star} =  d^{1\star} \frac{x}{1+x} - \frac{1}{1+x} \\
\label{fs1B}
& & f^{\star1} = d^{1\star} \frac{1}{1+x} - \frac{x}{1+x} \\
\label{f1s1B}
& & f^{1\star1} = - d^{1\star} \\
\label{f10sB}
& & f^{10\star} = d^{\star1}\frac{1}{1+x}  \\
\label{fs01B}
& & f^{\star01} = d^{\star1}\frac{x}{1+x} 
\eea
which must be satisfied for \eref{sd} to be stationary. The relations for minimal
interaction range follow by setting $f^{\star01} = f^{10\star} = d^{\star1} = 0$.
Here we only outline the strategy of the proof. More mathematical details are given in the appendix.

Dividing \eref{me1} by the stationary distribution \eref{sd}, the stationarity 
condition becomes
\bel{sc}
\sum_{i=1}^N \left[\omega_i(\bfeta_{tl}^{i}) \frac{\pi(\bfeta_{tl}^{i})}{\pi(\bfeta)}
+ \kappa_i(\bfeta_{rel}^{i}) \frac{\pi(\bfeta_{rel}^{i})}{\pi(\bfeta)} 
- (\omega_i(\bfeta)+\kappa_i(\bfeta)) \right] = 0.
\ee
Now we introduce the quantities
\bea
\label{Ddef}
D_i(\bfeta) & = & \omega_i(\bfeta_{tl}^{i}) \frac{\pi(\bfeta_{tl}^{i})}{\pi(\bfeta)} - \omega_i(\bfeta) \\
F_i(\bfeta) & = & \kappa_i(\bfeta_{rel}^{i}) \frac{\pi(\bfeta_{rel}^{i})}{\pi(\bfeta)} - \kappa_i(\bfeta).
\eea
Taking into account periodicity, the stationarity condition \eref{sc} is satisfied if
the {\it lattice divergence condition} 
\bel{ldc}
D_{i}(\bfeta) + F_{i}(\bfeta) = \Phi_{i}(\bfeta) - \Phi_{i+1}(\bfeta)
\ee
holds for all allowed configurations with a family of functions $\Phi_{i}(\bfeta)$ satisfying 
$\Phi_{N+1}(\bfeta) = \Phi_{1}(\bfeta)$. The
lattice divergence condition can be understood as a specific discrete form of Noethers theorem.

A further ingredient is to consider the {\it headway}, i.e., the number of empty sites $m_i$
between neighbouring rods $i$ and $i+1$, as random variable rather than the
actual positions $x_i$, which is possible due to translation invariance. A configuration of RNAPs
is then specified by the distance vector $\bfm := (m_1,\dots, m_N)$
and the state vector $\bfs$. The energy \eref{energy}
becomes $E(\bfm) = J \sum_i \delta_{m_i,0}$. As a result, the stationary distribution \eref{sd}
becomes a product measure in the new distance variables $m_i$. The rates $\omega_i$ 
and $\kappa_i$ are functions of the distances $m_{i-1}$ and $m_i$.
By somewhat lengthy, but straightforward computations one then finds that the 
lattice divergence condition is satisfied if and only if relations \eref{xB} - \eref{fs01B}
are satisfied. 

\subsection{Pure hard-core repulsion}

It is interesting to note that In the absence of a static nearest-neighbour interaction ($J=0$) other 
than hard-core repulsion one has $y=1$. Then the stationarity conditions \eref{xB} - \eref{fs01B} 
force all parameters that determine the interaction contributions to the translocation rate and to the 
PP\textsubscript{i} release rate to be zero. Thus a pure hard-core repulsion is compatible with transition
rates of the interacting system that are identical to the transition rates of a single RNAP.
Therefore, stationary of a distribution corresponding to hard-core repulsion only is \emph{not} 
compatible with any kinetic mechanism that produces the experimentally observed pushing phenomenon.

\subsection{Pushing and blocking}

As stressed above, the model assumptions include the form of the stationary distribution \eref{sd}
and some assumptions on the form of the transition rates, as expressed by the parameters
appearing in \eref{jrA} and \eref{rrA} for minimal kinematic interaction range and \eref{jrB} and 
\eref{rrB} resp. for the extended kinematic interaction range. There is no assumption on the 
magnitude of these parameters. The quantitative relations \eref{xB} - \eref{fs01B} are 
\emph{consequences} arising from requiring stationarity of the distribution \eref{sd}. 

We consider first the choice of minimal kinematic
interaction range. Inserting \eref{yB} with $d^{\star1}=0$ into the translocation rate \eref{jrA} one finds
the translocation rate
\bel{jrA1}
\omega_i(\bfeta) = \omega^{\star} \delta_{s_i,1} y^{\delta_{x_{i-1}+\ell , x_{i}}}
\left(1-\delta_{x_{i}+\ell, x_{i+1}}\right) .
\ee
This form of the rate means the following. If RNAP $i$ has no left trailing neighbour and the 
neighbouring site in forward
direction is free, then translocation takes place with the rate $\omega^{\star}$ of an isolated
RNAP. However, if a trailing RNAP has arrived to the left of the RNAP, then translocation 
takes place with rate 
\bel{pushrate}
\omega^{push} = y \omega^{\star}.
\ee
For repulsive static interaction ($J>0$) one has $y>1$. Thus the trailing RNAP pushes the
leading RNAP, as observed in experiments \cite{Epsh03a,Epsh03b}. Blocking, i.e., the
prevention of a forward step through the presence of an RNAP on the target site, occurs
in the same manner as in the case of pure hard-core repulsion, since the translocation
rate in the minimal interaction range scheme is not sensitive to the occupation of the
location 2 sites ahead of its present position.
The pushing which is discussed here in terms of the rate $\omega^{push}$
will from now on be referred to as \emph{stochastic pushing}, as opposed to a collective increase 
of the average RNAP velocity discussed below. 

We regard to the extended interaction range we point out that
according to \eref{yB}, repulsion corresponds to $d^{1\star} > d^{\star1}$.
The significance of this relation becomes clear by noting that according to the
definition \eref{jrB} the quantity $\omega^{\star} (1+d^{1\star})$ is the rate for moving away from
from a left-neighbouring trailing rod (without at the same time acquiring a leading rod
a new right neighbour), while $\omega^{\star} (1+d^{\star1})$ is the rate for approaching 
a right-neighbouring leading rod (without at the same time losing a trailing rod
as previous left neighbour). The case $d^{\star1} < 0$ will be referred to as \emph{stochastic 
blocking enhancement} or simply \emph{jamming}, since in this range of the interaction 
parameter $d^{\star1}$ the rate of approaching an
RNAP is reduced compared to the translocation rate of an single non-interacting RNAP.

\subsection{Average excess and dwell time}

The simplest measure that characterizes the distribution of RNAP is the average excess 
density $\sigma = (\exval{N^1} - \exval{N^2})/L$ of RNAP with no PP\textsubscript{i} bound
over the PP\textsubscript{i} bound state of RNAP. From the grandcanonical stationary distribution
\eref{sdgc} one obtains
\bel{statexcess}
\sigma = - \frac{k_B T}{L} \frac{\rmd}{\rmd \lambda} \ln Z_{gc} 
= \frac{1-x}{1+x} \rho .
\ee
For the densities of each RNAP state
\be 
\rho^\alpha := \exval{\delta_{s_i,\alpha}} = \frac{1}{L} \exval{N^\alpha} 
\ee
one concludes
\bel{statN}
\rho^1 = \frac{1}{1+x} \rho, \quad \rho^2 = \frac{x}{1+x} \rho .
\ee

Due to ergodicity this ensemble average is proportional to the average fraction of dwell 
times $\tau^\alpha = \rho^\alpha/\rho$ that the RNAP spends in state $1,2$. Thus
\bel{stattau}
\tau^1 = \frac{1}{1+x}, \quad \tau^2 = \frac{x}{1+x} .
\ee
It is noteworthy that the dwell times of an RNAP in the interacting ensemble exhibit no difference 
to the case of a single RNAP.
Using the result \eref{xB} that follows from the requirement of stationarity we arrive 
at the balance equation 
\be 
\frac{\rho^1}{\rho^2} = \frac{\tau^1}{\tau^2} = \frac{\omega^{\star}}{\kappa^{\star}} 
\ee
which expresses the ensemble ratio in terms of the single-RNAP translocation rate 
$\omega^{\star}$ and the single-RNAP PP\textsubscript{i} release rate $\kappa^{\star}$.

\subsection{Elongation rate}

Since at each translocation step the RNA transcript is elongated by one nucleotide, the average
speed $v$ of an RNAP in units of bp per second equals the elongation rate as defined 
by the number of polymerized nucleotides per second. In a collection of interacting RNAP a 
measure of the total elongation rate is the stationary RNAP 
flux $j$, which is defined as the average number of RNAP translocations
per second and basepair \cite{Trip08}. The average flux and the average speed are 
related by
\bel{jrv} 
j =  \rho v .
\ee
The average flux is the expectation of translocation rate $\omega_i(\bfeta)$ in the stationary 
distribution \eref{sd}.

In order to elucidate the effect of interactions on the average elongation rate we first consider a 
single isolated RNAP as a reference. From the work of Wang et al \cite{Wang98} one can compute the 
mean velocity 
\bel{sv}
v^{\star} = \frac{\omega^{\star} \kappa^{\star}}{\omega^{\star} + \kappa^{\star}} =
\frac{1}{1 + \frac{\omega^{\star}}{\kappa^{\star}}} \, \omega^{\star} .
\ee
The expression \eref{sv} differs from 
$v^{\star}_0=\omega^{\star}$ for a simple random walk 
by a prefactor of Michaelis-Menten form due to the need to go through the intermediate
state $2$. From the result \eref{stattau} for the mean dwell time it becomes clear
that this prefactor is the average time spent in the mobile state 1.
%

\subsubsection{Minimal interaction range}

For minimal interaction range the translocation rate is given by \eref{jrA}. Using the 
factorization property of the stationary distribution one gets
\be 
j = \omega^{\star} \rho^1 \exval{\left(1 + d^{1\star} \delta_{x_{i-1}+\ell , x_{i}} \right)
\left(1-\delta_{x_{i}+\ell, x_{i+1}}\right)}.
\ee
With the definition \eref{thetadef} from the appendix and the distribution \eref{sdgc}
one obtains
\bea
\label{vmir}
v & = &  A(\rho,y) v^{\star} 
\eea
with 
\be 
A(\rho,y) = y \frac{1-\ell\rho}{\rho} \tilde{z}(\rho,y) 
\ee
and the function 
\be 
\tilde{z}(\rho,y) = 1 - \frac{1-(\ell-1)\rho - \sqrt{(1-(\ell-1)\rho)^2 - 4 \rho (1-\ell\rho)(1-y^{-1}) }}{2 \rho (1-y^{-1})}
\ee
related through the mean headway to the function $z(\rho,y)$ given in \eref{zsol} 
\be 
\frac{1-\tilde{z}}{1-z} = \frac{1}{\rho} - \ell .
\ee

Comparing \eref{vmir} with the single-RNAP scenario one notices that speed and flux
change by a factor that is entirely determined by the RNAP density and the 
static interaction through the parameter $y$ \eref{xydef}. This factor does not
depend on the translocation rate 
$\omega^{\star}$ and the rate $\kappa^{\star}$ for PP\textsubscript{i} release
for a free RNAP.

\subsubsection{Extended interaction range}

For extended interaction range the translocation rate given by \eref{jrB} acquires a further
term. The factorization property of the stationary distribution yields
\be 
j = \omega^{\star} \rho^1 \exval{\left(1 + d^{1\star} \delta_{x_{i-1}+\ell , x_{i}} 
+ d^{\star1} \delta_{x_{i+1}, x_{i}+\ell + 1} \right) \left(1-\delta_{x_{i}+\ell, x_{i+1}}\right)}.
\ee
From the partition function \eref{sdgc} computed in the Appendix one finds
\bea
\label{veir}
v & = &  v^{\star} B(\rho,y,d^{\star1}) 
\eea
where
\be
B(\rho,y,d^{\star1}) = \frac{d^{1\star} (1-\ell \rho) \tilde{z}(\rho,y)
- d^{\star1} \rho z(\rho,y)}{\rho( 1-y^{-1})} 
\ee
and $z(\rho,y)$ defined in \eref{zsol}.
In contrast to the previous case, the average RNAP speed and average elongation rate cannot 
be expressed in terms of the
interaction parameter $y$ alone, but are functions of both interaction parameters $d^{1\star}$
and $d^{\star1}$. As pointed out above,  $d^{1\star}>d^{\star1}$ corresponds to static repulsion.

\subsection{Relation to experiments and predictions}

The model makes predictions about the kinetics of transcription elongation and, specifically,
about the rate of elongation and total rate of RNA synthesis in dependence on the rate of translocation, 
and pyrophosphate release on the one hand, and on the density of and interaction strength between RNAP 
initiating from the same promoter sequence on the other hand. In all plots shown below we use the parameter 
value $\ell=5$.

\begin{figure}[h]
\begin{center}
\includegraphics*[width=0.5\textwidth]{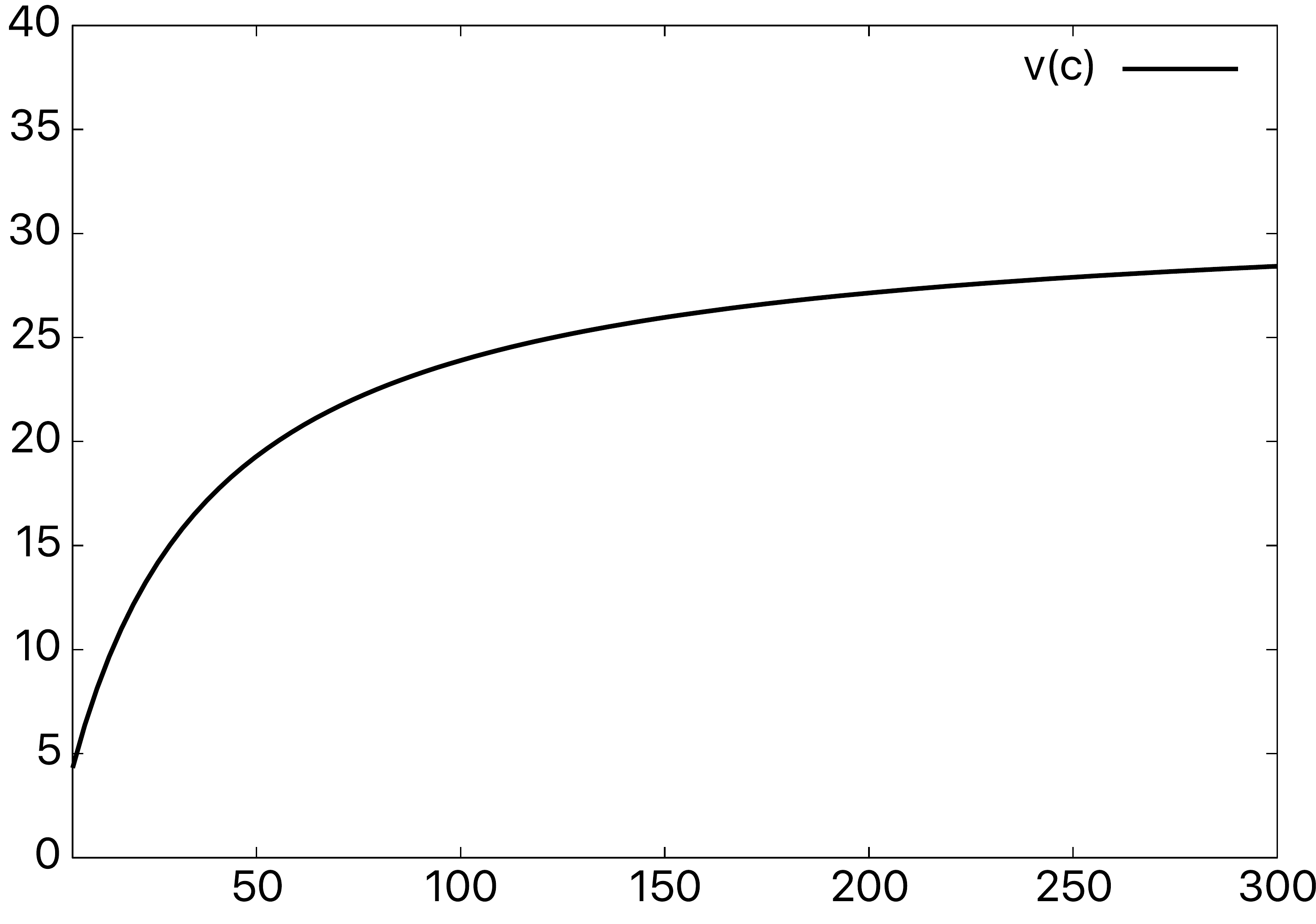}
\end{center}
\caption{\label{Fig:RNAP_P1_Fig2}
Average RNAP velocity $v(c)$ in $bp/s$ as function of NTP concentration $c$ in $\mu M$
at amplitude value $A(\rho,y)=1$.}
\end{figure}

The dependence of the elongation rate on the translocation rate is experimentally directly accessible by 
manipulating the NTP concentration. The model predicts that this dependence on NTP concentration is 
independent of RNAP density and interactions, except for the overall amplitude. \Fref{Fig:RNAP_P1_Fig2} shows the 
average RNAP velocity as a function of NTP concentration $c = [NTP] (\mu M)^{-1}$ with 
amplitude chosen as $A(\rho,y)=1$ (see below).
We have taken as rate of \PPi release the value $\kappa^{\star}=31.4s^{-1}$ obtained in \cite{Wang98}. 
Notice that only the ratio of NTP concentration and \PPi release rate enter the analytical expression. Hence 
the model predicts the same curve for a different rate \PPi release rate $\tilde{\kappa} = a \kappa^{\star}$ 
if one works with concentration $\tilde{c} = a c$.

\begin{figure}[h]
\begin{center}
\includegraphics*[width=0.45\textwidth]{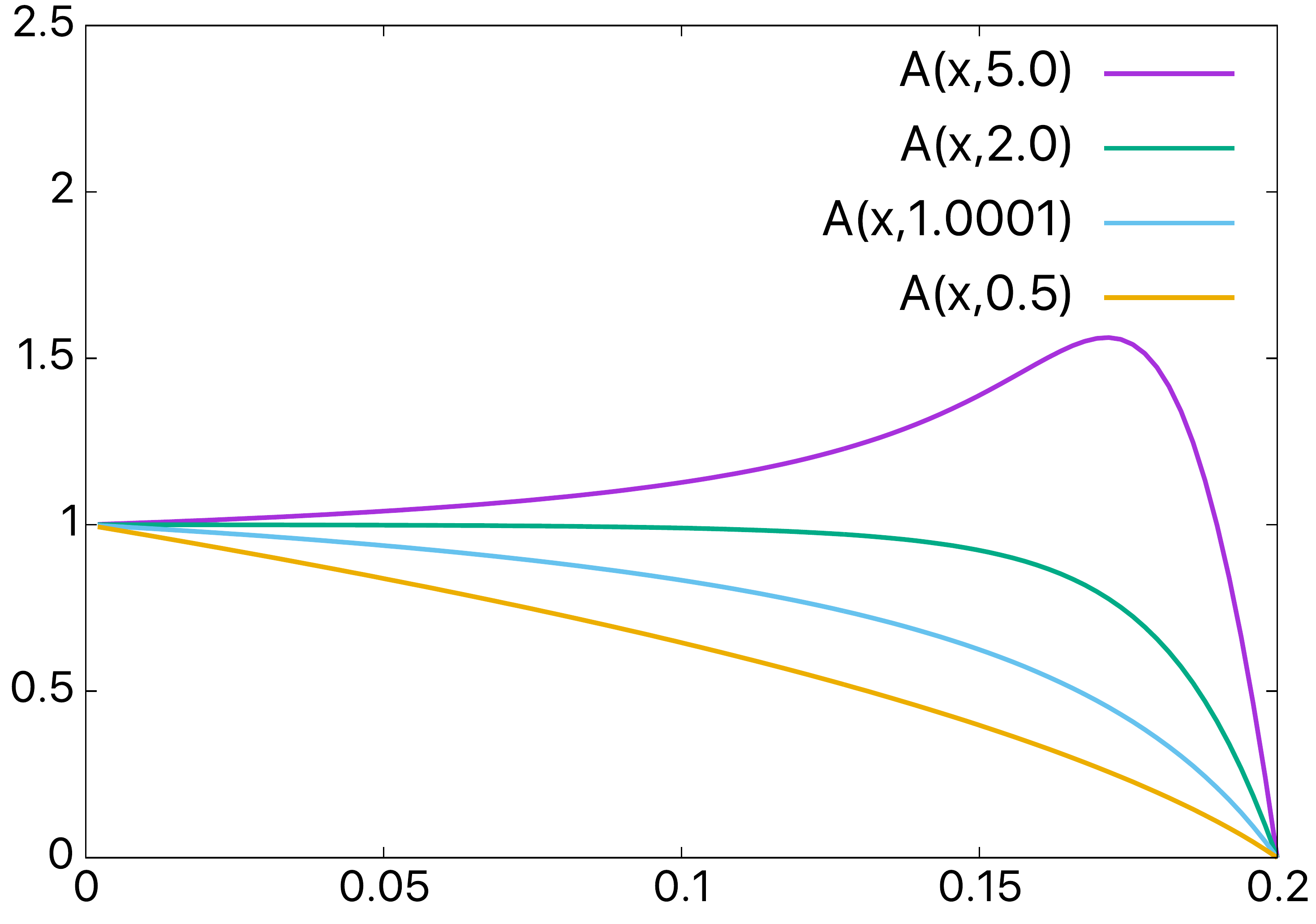}
\includegraphics*[width=0.45\textwidth]{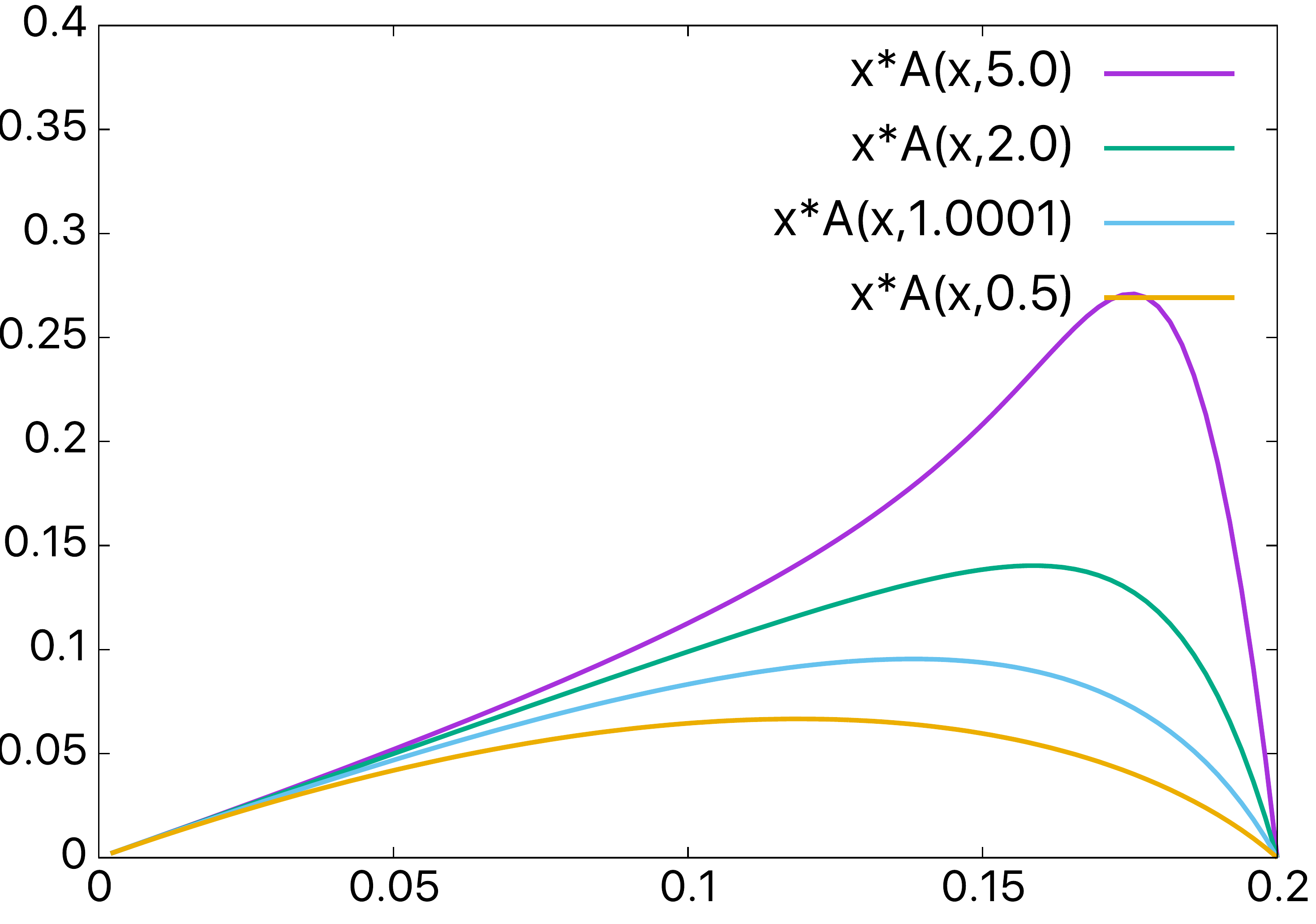}
\end{center}
\caption{\label{Fig:RNAP_P1_Fig3}
RNAP velocity amplitude $A(\rho,y)$ (left panel) and
RNAP flux amplitude $\rho A(\rho,y)$ (right panel) as function of RNAP density for different interaction 
strengths $y$ and minimal interaction range. 
Curves from top to bottom: $y=5.0$ (strong repulsion), 
$y=2.0$ (critical repulsion strength), $y=1.0001$ (only hard core), $y=0.5$ (attraction).
The dotted reference line corresponds to non-interacting RNAP.}
\end{figure}

By using an excess of DNA one can regulate the number of RNAP initiating from a single DNA 
molecule and thus the RNAP density. We consider first the predictions for minimal interaction range.
As a function of the density, the speed increases and reaches maximum at $\rho^\ast>0$ only 
if $y>2$. This can be shown analytically by taking the derivative of \eref{vmir} w.r.t. the density.
Thus, while stochastic pushing always occurs in single jumps, it has a cooperative effect on the 
avarage speed of a single RNAP in an ensemble of interacting RNAP 
only for sufficiently strong repulsion, up to some characteristic blocking density $\rho_b$ above which blocking
dominates and the velocity drops below that of an isolated RNAP without pushing. 
Weak short-range repulsion below the critical value (or attraction) reduce the speed of the
RNAP as the density increases, as does pure hard-core repulsion. Hence this regime is 
blocking-dominated, i.e., the pushing effect is not strong enough to compensate the
blocking due to steric hard-core repulsion. These features are shown in Fig. \ref{Fig:RNAP_P1_Fig3}, left panel.

In contrast, the total average elongation rate $j=\rho v$ increases for any interaction
strength, even for attractive interaction. It reaches a maximum at a second critical density 
$\rho^{\ast\ast} > \rho^\ast$ (Fig. \ref{Fig:RNAP_P1_Fig3}, right panel) 
where for $y\leq 2$ we define $\rho^\ast=0$. However, cooperativity sets in again only
above the critical repulsion strength $y_c=2$. Otherwise the flux of interacting RNAP is lower than
the flux generated by the same number of single RNAP (dotted curve in the right panel).
We stress that pushing is not an input in our calculation of the transition rates, 
but is a prediction borne out by experiments.

Next we point out our predictions for situations where experimental results are currently not available.
We consider manipulating the kinetic interaction range which may be probed by applying an 
external torque to the RNAP, as has been suggested as a means to to regulate the average velocity of 
RNAP \cite{Trip09b}. Taking in \eref{veir} the derivative w.r.t. the density at density 0 
one finds that the average speed increases with density provided that
\bel{limcurve} 
d^{1\star} > - 1/2 + y^{-1}.
\ee
Since $d^{\star1} < 0$ (stochastic blocking enhancement) is assumed, cooperative pushing at low densities
requires the same minimal interaction strength $y_c=2$ as in the case of minimal interaction range.

\begin{figure}[h]
\begin{center}
\includegraphics*[width=0.45\textwidth]{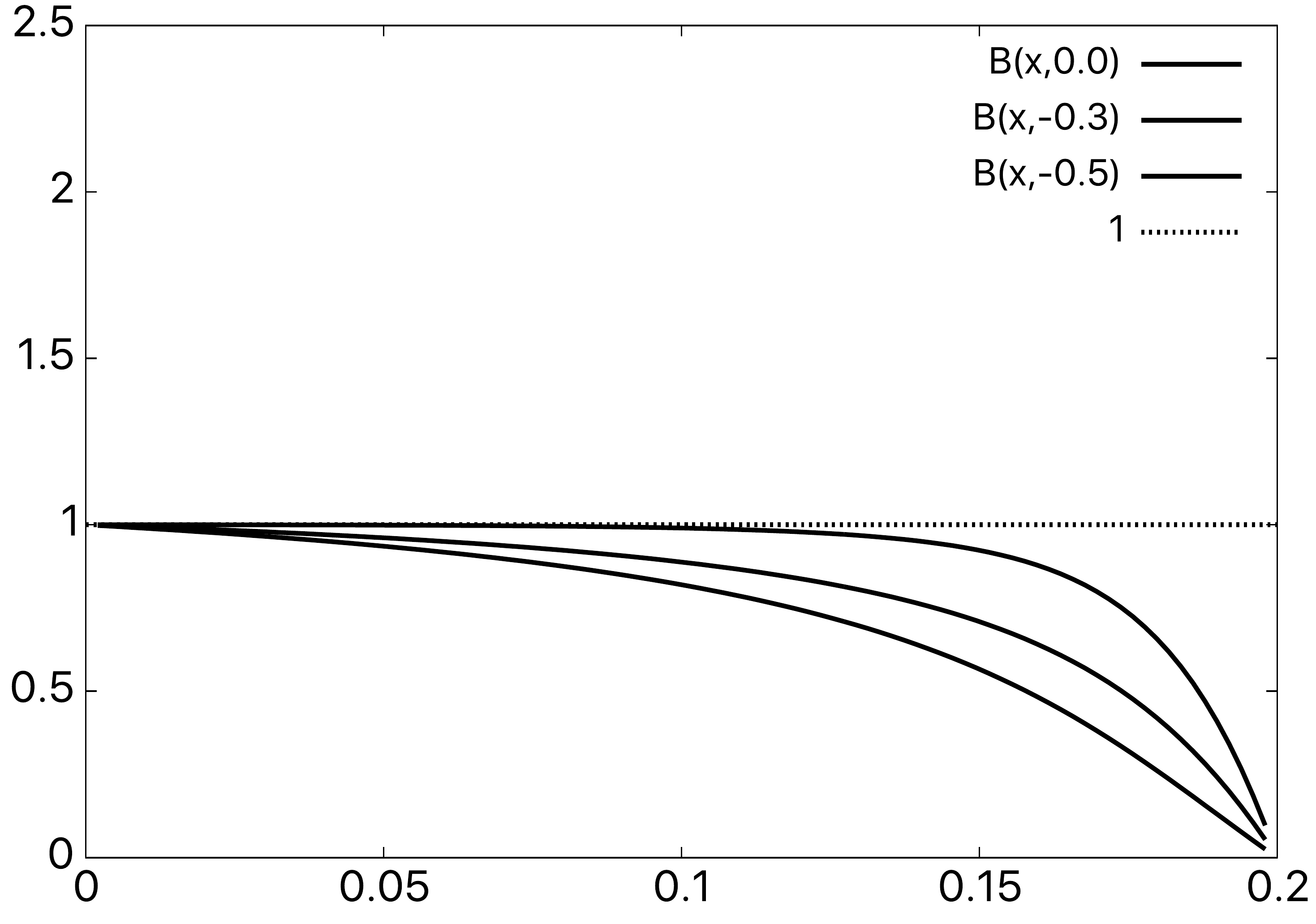}
\includegraphics*[width=0.45\textwidth]{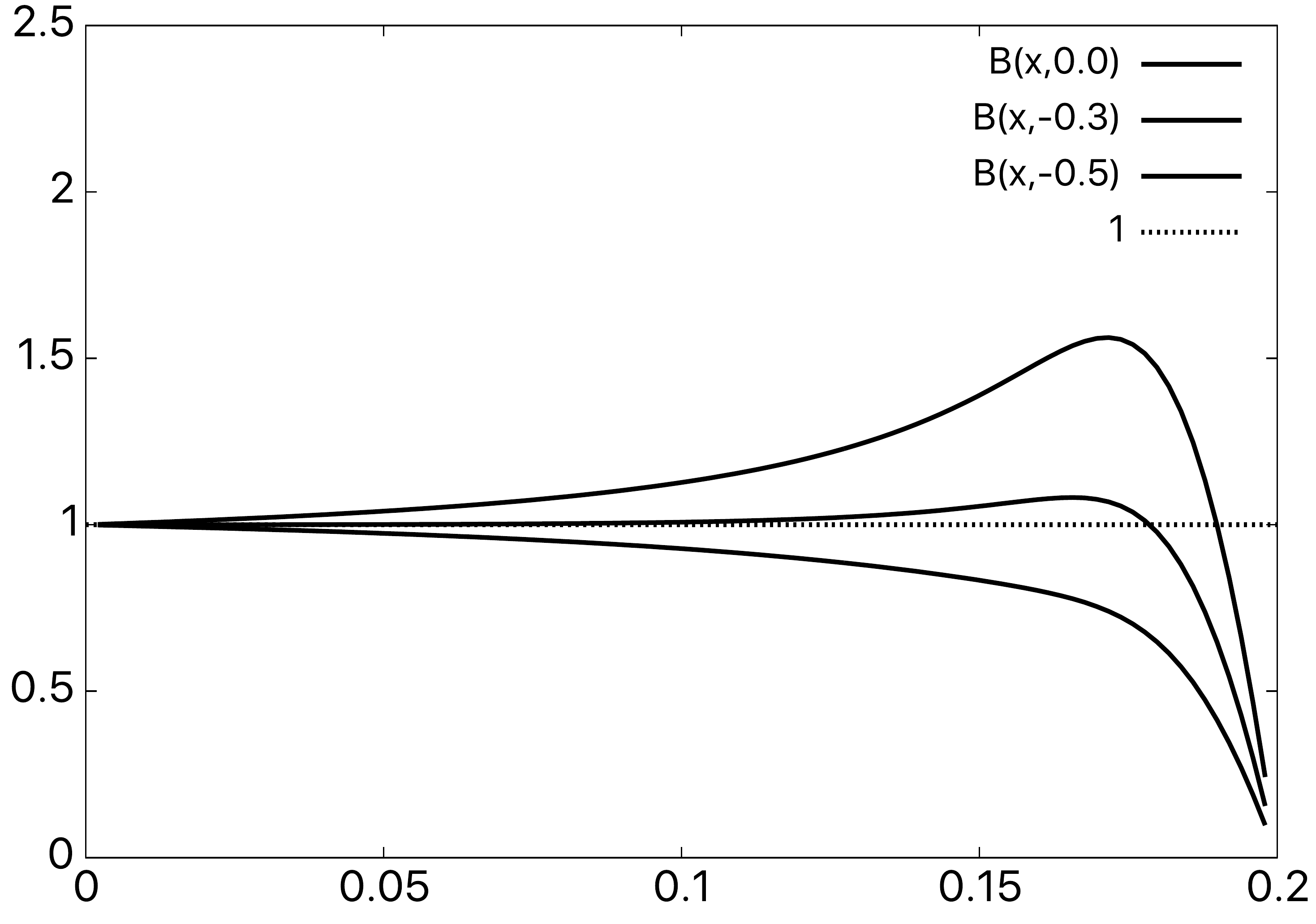}
\end{center}
\caption{\label{Fig:RNAP_P1_Fig5}
RNAP velocity amplitude $B(\rho,y,d^{\star1})$ for interaction strengths $y=2$ (left panel) and 
$y=5$ (right panel) as function of 
RNAP density for different values of $d^{\star1}$ that characterize the extended interaction range. 
Curves from top to bottom: $d^{\star1}=0$ (reference curve without extension of interaction range), 
$d^{\star1}=-0.3$ (intermediate blocking enhancement), $d^{\star1}=-0.5$ (critical blocking enhancement).
The dotted reference line corresponds to non-interacting RNAP.}
\end{figure}

In terms of the interaction parameters $d^{1\star},d^{\star1} > -1$ the limiting curve 
\eref{limcurve} is in the domain $d^{\star1} > -1/2$, $d^{1\star}>0$. 
This has a twofold meaning. First, stochastic pushing ($d^{1\star}>0$) is necessary for cooperative 
pushing to emerge. When the stochastic blocking enhancement
is too strong ($d^{\star1} \leq d^{\star1}_c = -1/2$), then even strong stochastic 
pushing ($y$ arbitrarily large) does not lead to \emph{cooperative} pushing at low densities.
In Fig. \ref{Fig:RNAP_P1_Fig5} these effects are illustrated by for $y=y_c$ (left panel, no cooperative pushing
at any density) and for strong repulsion $y=5$ (right panel, cooperative pushing until jamming takes over
at high density, provided stochastic blocking enhancement is not too strong ($d^{1\star} > 0.3$).

\begin{figure}
\begin{center}
\includegraphics*[width=0.45\textwidth]{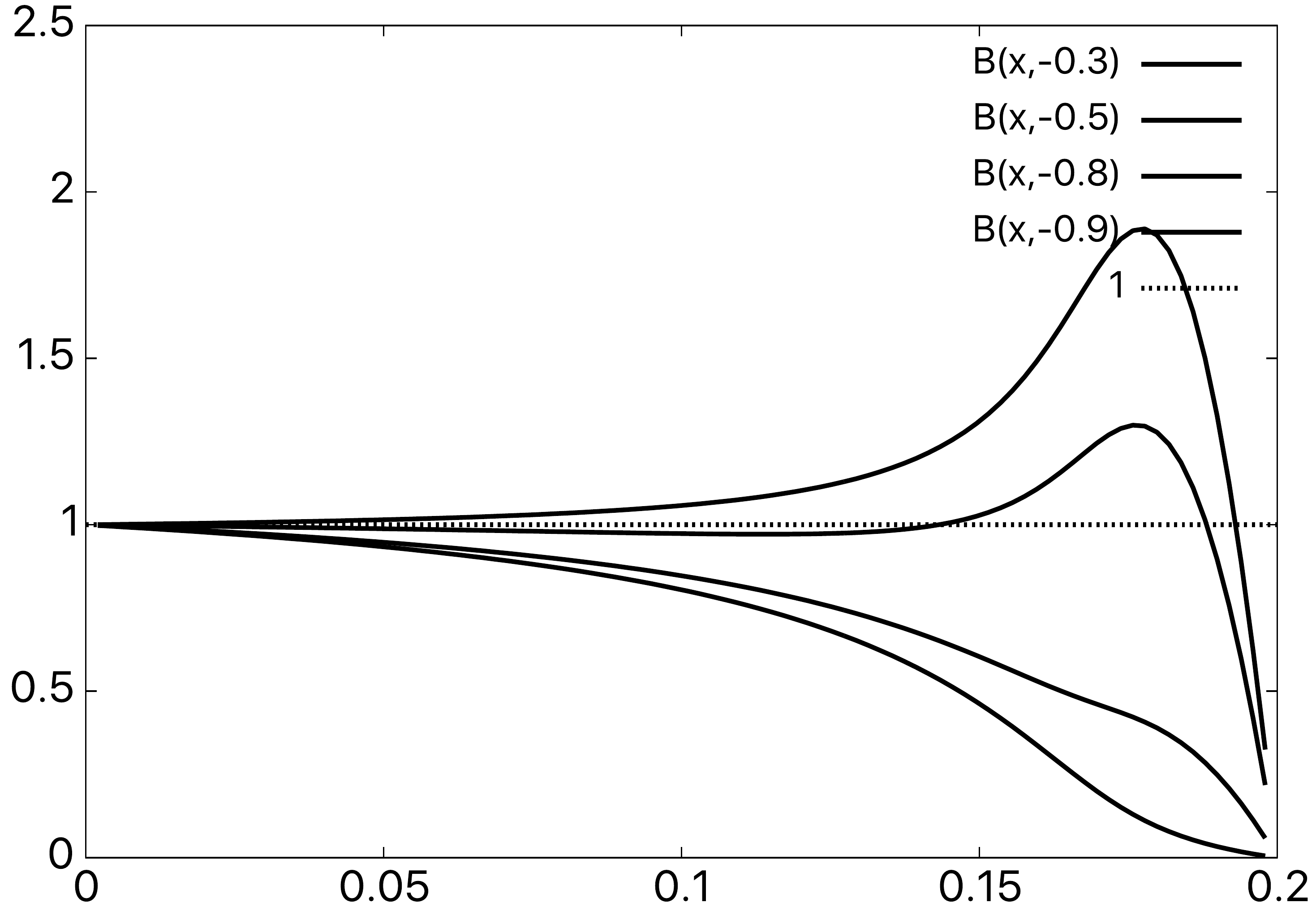}
\includegraphics*[width=0.45\textwidth]{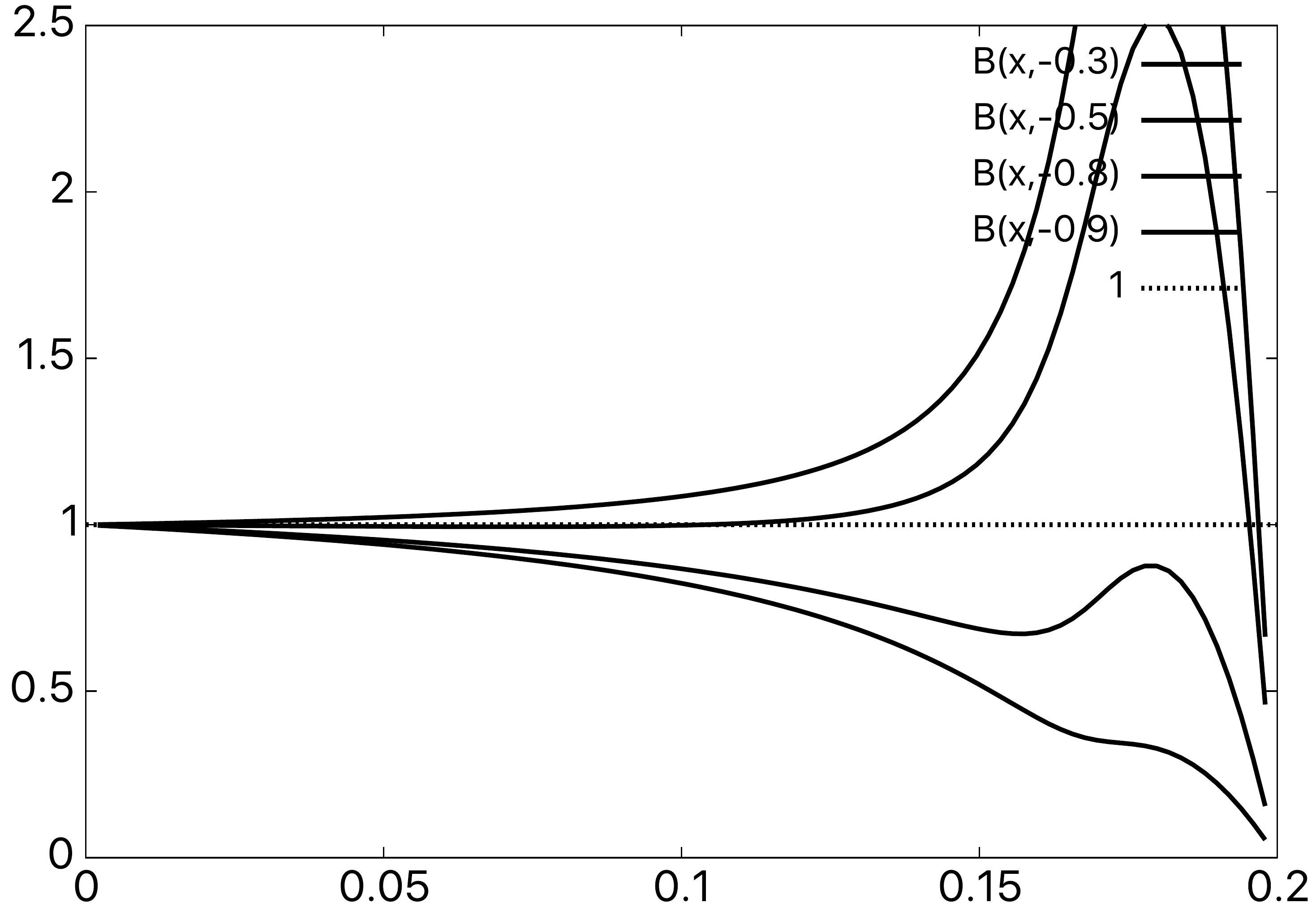}
\end{center}
\caption{\label{Fig:RNAP_P1_Fig6}
RNAP velocity amplitude $B(\rho,y,d^{\star1})$ for large interaction strengths $y=10$ (left panel) and 
$y=20$ (right panel) as function of 
RNAP density for different values of $d^{\star1}$. 
Curves from top to bottom: $d^{\star1}=-0.3,-0.5,-0.8,-0.9$.
The dotted reference line corresponds to non-interacting RNAP.}
\end{figure}

\begin{figure}
\begin{center}
\includegraphics*[width=0.45\textwidth]{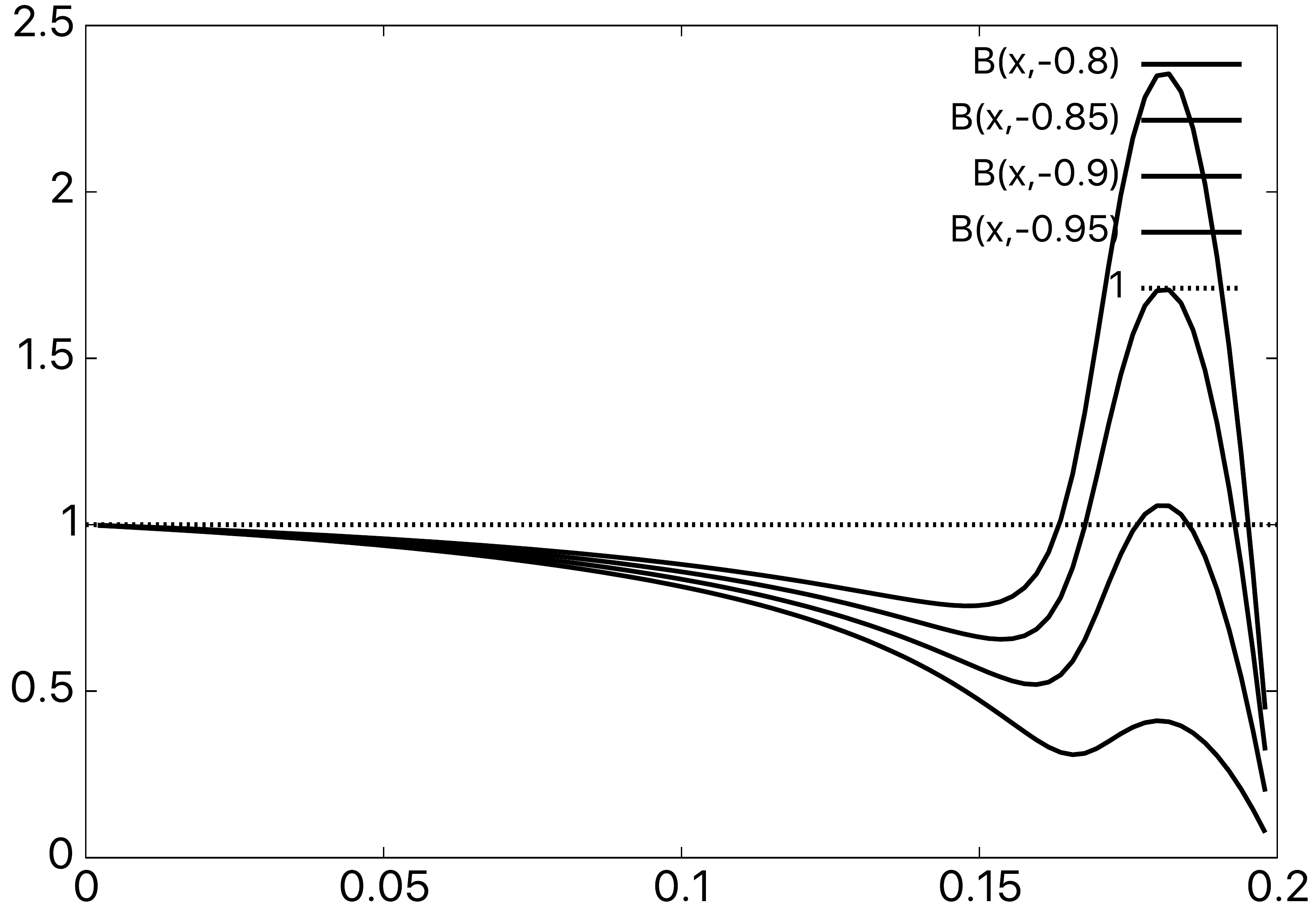}
\includegraphics*[width=0.45\textwidth]{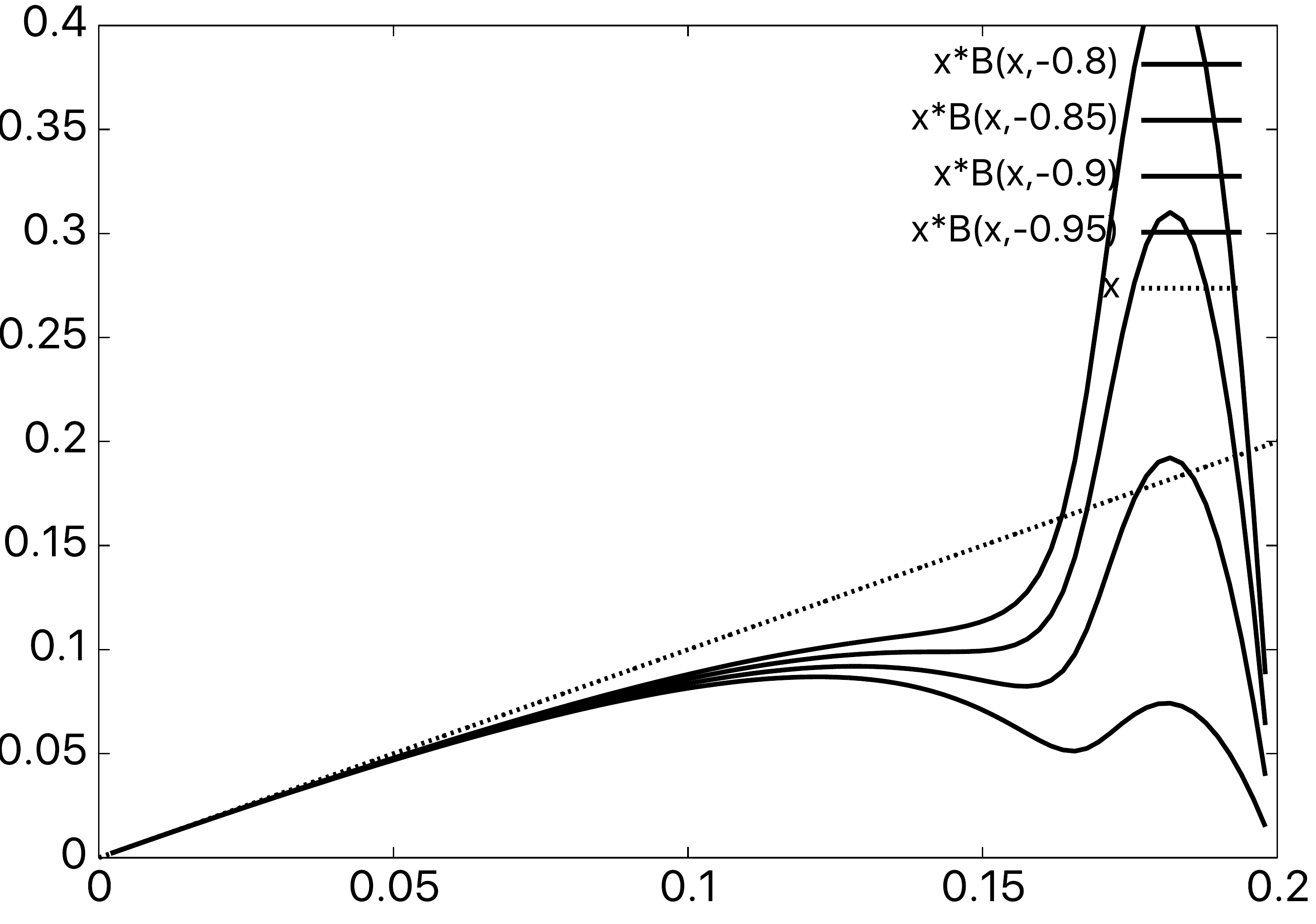}
\end{center}
\caption{\label{Fig:RNAP_P1_Fig7}
RNAP velocity amplitude $B(\rho,y,d^{\star1})$ (left panel) and flux amplitude (right panel)
for large interaction strength $y=50$ as function of 
RNAP density for different values of $d^{\star1}$. 
Curves from top to bottom: $d^{\star1}=-0.8,-0.85,-0.9,-0.95$.
The dotted reference line corresponds to non-interacting RNAP.}
\end{figure}

When both repulsive static interaction and stochastic blocking enhancement are very strong, a new 
phenomenon appears. The speed of an RNAP develops an intermediate minimum at a density $\rho_m$. 
The local minimum deepens with increasing static repulsion and is outside the range of
cooperative pushing, even though there is stochastic pushing. Somewhat paradoxically,
however, at higher densities there a reentrance to a cooperative pushing regime, see 
Fig.~\ref{Fig:RNAP_P1_Fig6} for two different values of $y$. 
For a range of densities stochastic pushing prevails over stochastic blocking enhancement. Only at
very high densities cooperative pushing disappears again. 
The drop is also present in the average elongation rate at a higher density, see 
Fig.~\ref{Fig:RNAP_P1_Fig7} where the velocity amplitude $B(\rho,y,d^{\star1})$ (left panel) and the flux 
amplitude $\rho B(\rho,y,d^{\star1})$ (right panel) are shown for a large value $y=50$ of the static repulsion 
strength. The large value was chosen to demonstrate that the reentrance effect can be
very pronounced and is very sensitive to the interaction parameter $d^{\star1}$.

\section{Conclusions}

The elongation phase can involve multiple RNAPs moving one after 
another along the same DNA molecule. Therefore an investigation of the kinetics of transcription requires
taking into account interactions between RNAP. Experimental results that demonstrate a pushing effect
of trailing RNAP necessitate a description of such interactions that goes beyond a steric hard-core repulsion
through additional short-range interactions. To our knowledge the approach developed in this paper 
is the first attempt to capture such extra interactions as well as the mechanochemical cycles of each 
individual RNAP in the elongation stage in a stochastic model that allows for an 
exact analytical computation of its stationary properties. In particular, we make precise 
predictions on the average rate of elongation 
as a function of interaction strength as well as the RNAP density and NTP concentration which can be
manipulated experimentally so that model predictions can be tested.

The model is based on a reduced picture
of the mechanochemical cycle and incorporates a nearest-neighbour interaction
that allows us to understand certain qualitative and quantitative aspects of the kinetics of transcription
and the dependence of the rate of elongation on interactions. 
Without using the experimental results on pushing in the computation of the
transition rates, the model \emph{predicts} pushing: When a trailing RNAP reaches a paused RNAP,
then the paused RNAP continues stochastically with a higher translocation rate as in the absence of the
trailing RNAP. 
A stationary distribution for RNAP with hard-core repulsion only is \emph{not} 
compatible with any kinetic mechanism that produces the experimentally observed 
pushing phenomenon. This mechanism requires an RNAP distribution
that incorporates an additional static short-range repulsion.

However, according to the model, the \emph{cooperative}
effect of individual RNAP pushing, i.e, the enhancement of the average elongation rate, requires
an effective short-range repulsion between RNAP beyond a certain strength. Given this 
interaction strength, cooperative pushing persists up to
a maximal RNAP density beyond which jamming due to hard-core repulsion prevails. At such
high RNAP densities the elongation rate of multiple RNAP transcribing on the same
DNA segment is reduced compared to the same number of non-interacting RNAP, i.e.,
an average over RNAP transcribing the same RNA independently on different DNA molecules.

A second outcome of our approach is a certain degree of freedom in the choice
of kinetic processes that do not alter the stationary distribution. In particular, one can
describe not only pushing, but also a stochastic blocking enhancement that goes beyond
the effects of pure hard-core repulsion by a reduction of the translocation rate when a trailing
RNAP approaches a paused RNAP. At present there are not 
sufficient experimental data for any specific RNAP to determine these parameters. For the sake 
of simplicity, we have illustrated the effect of the interplay of stochastic pushing 
and stochastic blocking enhancement for the simplest case compatible
with stationarity of the RNAP distribution. 

It turns out that the RNAP velocity and transcription rate 
are sensitive to this interplay in a subtle and somewhat surprising manner.
At low densities and moderately strong repulsion cooperative pushing occurs 
much in the same way as without stochastic blocking enhancement, provided 
that stochastic blocking enhancement is not too strong. When stochastic blocking enhancement
is very strong, then a moderate stochastic pushing is not sufficient to generate a cooperative
pushing effect. However, when also the repulsion (and therefore stochastic pushing) is very strong
then as RNAP density increase, the elongation rate first decreases (due to jamming), but it reaches
a minimum and then enters a range of RNAP density, where cooperative pushing reappears.
Only at very high RNAP density jamming dominates again.

Thus the model provides an explanation of cooperative pushing in terms
of static and kinetic interactions between neighbouring RNAP.
The modelling approach is mathematically robust and can be extended to allow for a more
detailed biological description of the mechanochemical cycle of the RNAP during elongation, for back-tracking,
and also for incorporating more general short-range interactions.


\section{Acknowledgements}

G.M.S. thanks the Instituto de Matem\'atica e Estat\'astica of the University of S\~ao Paulo 
for kind hospitality. This work was financed in part by Coordenação de Aperfeiçoamento de 
Pessoal de Nível Superior -- Brazil (CAPES) -- Finance Code 001,  by the grants 
2017/20696-0, 2017/10555-0 of S\~ao Paulo Research Foundation (FAPESP), and by the grant 309239/2017-6 of
Conselho Nacional de Desenvolvimento Cinetífico e Tecnológico (CNPq). The authors thank D. Chowdhury (IIT Kanpur) for useful comments
on an early draft of this paper.

\appendix

\section{Mapping to the headway process}

We consider the extended interaction range. All results for minimal
interaction range follow by taking $d^{\star1}=f^{10\star}=f^{\star01}=0$.

\subsection{Master equation and stationary distribution}

Because of translation invariance one may describe the RNAP model in terms of the headway
distances $\bfm = (m_1,\dots,m_N)$ with $m_i = x_{i+1} - (x_{i}+\ell) \bmod L$ which is the 
number of vacant sites between the left edge of rod $i+1$ and the right edge of rod $i$. 
The total number of vacant sites is $M= L-N$. It is convenient to introduce the indicator
functions
\bel{thetadef} 
\theta_i^p := \delta_{m_i,p} = \delta_{x_{i+1},x_{i}+\ell+p}
\ee 
on a headway of length $p$ (in units of bp)
with the index $i$ taken modulo $N$, i.e., $\theta_0^p \equiv \theta_N^p$.
Since $\theta_i^p$ takes only values 0 or 1 one has
\bel{thetaexp}
y^{\theta_i^p} = 1 + (y-1) \theta_i^p
\ee
which is useful for explicit computations.

A translocation of RNAP $i$ corresponds to the transition $$(m_{i-1},m_i) \to (m_{i-1}+1,m_i-1)$$
with all other distances remaining unchanged. Due to steric hard-core repulsion this transition can only take
place if $m_{i} > 0$. In terms of the new stochastic variables $\bzeta = (\bfm,\bfs)$ given by the distance
vector $\bfm$ and the state vector $\bfs$ the transition rates \eref{jrB} and
\eref{rrB} for extended interaction range become
\bea 
\tilde{\omega}_i(\bzeta) & = & \omega^{\star} \delta_{s_i,1} \left(1-\theta^0_{i}\right) 
\left(1 + d^{1\star} \theta^0_{i-1} + d^{\star1} \theta^1_{i}\right) \\
\tilde{\kappa}_{i}(\bzeta) & = & \kappa^{\star} \delta_{s_i,2} \left(1 + f^{1\star} \theta^0_{i-1}
+ f^{\star1} \theta^0_{i}  + f^{1\star1} \theta^0_{i-1} \theta^0_{i} + f^{10\star} \theta^1_{i-1} 
+ f^{\star01} \theta^1_{i}\right) .
\eea
In the mapping to the headway process the stationary average speed of an RNAP 
is given by the stationary expectation of the function $\tilde{\omega}_i(\bzeta)$.

In order to write the master equation we need to introduce notation for the configurations that
lead to a given configuration $\bzeta$, viz. $\bzeta^{i-1,i}$ for translocation and $\bzeta^{i}$
for PP\textsubscript{i} release. For a fixed $\bzeta$ these configurations are defined by
\bea
& & \bfs^{i-1,i}_j = \bfs_j + (3-2s_j) \delta_{j,i}, \quad  \bfm^{i-1,i}_j = m_j + \delta_{j,i} - \delta_{j,i-1} \\
& & \bfs^{i}_j = \bfs_j + (3-2s_j) \delta_{j,i} , \quad  \bfm^{i}_j = m_j .
\eea
This yields the master equation
\bel{me2} 
\ddt P(\bzeta,t) = \sum_{i=1}^N Q_i(\bzeta,t)
\ee
with
\bea
Q_i(\bzeta,t) & = &
\tilde{\omega}_i(\bzeta^{i-1,i}) P(\bzeta^{i-1,i},t) - \tilde{\omega}_i(\bzeta) P(\bzeta,t)  
\nonumber \\
& & + \tilde{\kappa}_i(\bzeta^{i}) P(\bzeta^{i},t) - \tilde{\kappa}_i(\bzeta) P(\bzeta,t) .
\eea
This is the master equation for a misanthrope process \cite{Coco85} generalized to sites $i$ that
can take two degrees of freedom $s_i$.

An important property of the stationary distribution \eref{sd} is the independence of the probability
to find an RNAP in state 1 or 2 from the spatial distribution of RNAPs. An immediate consequence
is the factorization of the partition function \eref{pf} into into a summation over states and
a summation over the spatial degrees of freedom. Thus one finds 
for the stationary distribution in terms of the parameters \eref{xydef} and the new distance
variables \eref{thetadef} the expression
\be 
\tilde{\pi}(\bzeta) = \frac{1}{Z} \prod_{i=1}^N \left( y^{-\theta_i^0} x^{-3/2+s_i}\right)
\ee
which is of factorized form, indicating the absence of distance correlations.

\subsection{Proof of stationarity}

Notice that in the stationary state the master equation \eref{me2} has the form of a 
conservation law $\dot{Q} = 0$ where $Q=\sum_i Q_i$ with a
locally conserved stationary quantity $Q_i$. Invoking the discrete form of the Noether theorem
leads us to conclude that there is a locally conserved current $J_i$ associated with $Q_i$ such that
one can write $Q_i = J_{i-1} - J_i$. Defining
\bea 
\tilde{D}_i(\bzeta)  & = &
\tilde{\omega}_i(\bzeta^{i-1,i}) \frac{\tilde{\pi}(\bzeta^{i-1,i})}{\tilde{\pi}(\bzeta)} - \tilde{\omega}_i(\bzeta) \\
\tilde{F}_i(\bzeta)  & = & \tilde{\kappa}_i(\bzeta^{i}) \frac{\tilde{\pi}(\bzeta^{i})}{\tilde{\pi}(\bzeta)} - \tilde{\kappa}_i(\bzeta)  
\eea
one gets $Q_i = (\tilde{D}_i+\tilde{F}_i) \pi^{-1}$.
We can therefore rephrase the stationarity condition in a local divergence form equivalent to \eref{ldc}
with a locally conserved current $J_i=\tilde{\Phi}_i \pi$.

One has
\be
\theta_j^p(\bzeta^{i-1,i}) = \delta_{m_j + \delta_{j,i} - \delta_{j,i-1},p} = \theta_j^{p - \delta_{j,i} + \delta_{j,i-1}}(\bzeta), \quad
\delta_{s^i_i,\alpha} = \delta_{2s_i, 3-\alpha}
\ee
which yields
\bea 
\frac{\tilde{\pi}(\bzeta^{i-1,i})}{\tilde{\pi}(\bzeta)} 
& = & x^{3-2 s_i} y^{\theta_{i-1}^0 + \theta_{i}^0 - \theta_{i-1}^{1}} \\
\tilde{\omega}_i(\bzeta^{i-1,i}) 
& = & \omega^{\star} \delta_{s_i,2} \left(1-\theta^0_{i-1}\right)
\left(1 + d^{1\star} \theta^1_{i-1} + d^{\star1} \theta^0_{i}\right) 
\eea
and therefore
\bea
\label{DB2}
\tilde{D}_i(\bzeta)  
& = & x^{-1} \omega^{\star} \delta_{s_i,2} 
\left(1-\theta^0_{i-1}\right)
\left(1 + d^{1\star} \theta^1_{i-1} + d^{\star1} \theta^0_{i}\right)
y^{\theta_{i}^0 - \theta_{i-1}^{1}} \nonumber \\
& & - \omega^{\star} \delta_{s_i,1} \left(1-\theta^0_{i}\right) 
\left(1 + d^{1\star} \theta^0_{i-1} + d^{\star1} \theta^1_{i}\right) \\
\label{FB2}
\tilde{F}_i(\bzeta)  & = & \left( x \delta_{s_i,1} - \delta_{s_i,2} \right) \nonumber \\
& & \times \kappa^{\star} \left(1 + f^{1\star} \theta^0_{i-1}
+ f^{\star1} \theta^0_{i}  + f^{1\star1} \theta^0_{i-1} \theta^0_{i} + f^{10\star} \theta^1_{i-1} 
+ f^{\star01} \theta^1_{i}\right)
\eea

One notices that $\tilde{D}_i$ and $\tilde{F}_i$ depend only on the state at site $i$ and
only on the distance variables $\theta^0_{i-1}$, $\theta^1_{i-1}$, $\theta^0_{i}$, and $\theta^1_{i}$. 
This implies that the local divergence conditions holds \eref{ldc} if and only if
$$\tilde{D}_i + \tilde{F}_i = \tilde{\Phi}_{i-1} - \tilde{\Phi}_{i}$$ where
the quantity $\tilde{\Phi}_i$ is of the form $\tilde{\Phi}_i = a + b \theta^0_{i} + c \theta^1_{i}$ with
arbitrary constants $a,b,c$. This fact imposes conditions on the interaction parameters appearing in
\eref{DB2} and \eref{FB2} which are obtained by requiring 
$$\tilde{D}_i + \tilde{F}_i = [(a + b \theta^0_{i-1} + c \theta^1_{i-1})
- (a + b \theta^0_{i} + c \theta^1_{i})](\delta_{s_i,1}+\delta_{s_i,2}).$$
Comparing in \eref{DB2} and \eref{FB2} the
constant terms (independent of $\theta^\alpha_{i-1}$ and $\theta^\alpha_{i}$) one finds $a=0$ and
\eref{xB}. 
Comparing all the remaining terms proportional to $\delta_{s_i,1}$ one finds the relation \eref{f1s1B} and
\bea
& & f^{1\star} + f^{\star1} = d^{1\star} - 1 \\
& & f^{10\star} + f^{\star01} = d^{\star1}
\eea
corresponding to $b=f^{1\star}-d^{1\star}$ and $c=f^{10\star}$. Finally, using these preliminary results and
comparing each term in \eref{DB2} and \eref{FB2} proportional to $\delta_{s_i,2}$ to the corresponding term 
proportional to $\delta_{s_i,1}$ by using the indicator property \eref{thetaexp}
one arrives after some straightforward computation
at the relations \eref{yB}, \eref{f1sB}, \eref{fs1B}, \eref{f10sB}, and \eref{fs01B}.
This completes the proof of stationarity. 

\subsection{Partition function}

It is convenient to work in the grandcanonical ensemble 
defined by
\bel{sdgc}
\tilde{\pi}_{gc}(\bzeta) = \frac{1}{Z_{gc}} \prod_{i=1}^N \left( z^{m_i} y^{-\theta_i^0} x^{-3/2+s_i}\right)
\ee
where $Z_{gc} = (Z_1 Z_2)^N$ with
\be 
Z_1 = \frac{1+(y-1)z}{1-z}, \quad Z_2 = x^{\half} + x^{-\half}.
\ee

From this grandcanonical distribution one obtains the mean headway 
\bel{mhw}
\exval{m_i} = z \frac{\rmd}{\rmd z} \ln Z_{gc} = \frac{y z}{(1-z)(1+(y-1)z)}.
\ee
On the other hand, this is the mean available empty space on the lattice $L-\ell N$ per rod
which gives $\exval{m_i} = \frac{1}{\rho} - \ell$ where
\bel{rhodef}
\rho = \frac{N}{L}
\ee
is the RNAP density. Hence the auxiliary variable
$z$ is the solution of the quadratic equation
\bel{qez}
(y-1) z^2 + z \left( \frac{y \rho}{1-\ell\rho} - y + 2\right) -  1 = 0
\ee
given by
\bel{zsol}
z(\rho,y) 
= 1 - \frac{1-(\ell-1)\rho - \sqrt{(1-(\ell-1)\rho)^2 - 4 \rho (1-\ell\rho)(1-y^{-1}) }}
{2 (1-\ell\rho) (1-y^{-1})}.
\ee
Notice that the model allows for a number density in the range $0 < \rho \leq 1/\ell$.
This ensures that $0 \leq z < 1$ and therefore the partition function \eref{sdgc} is well-defined.


\end{document}